\documentclass[ preprint,  amsmath, amssymb,  showpacs, showkeys,aps, ]{revtex4-1}%
\usepackage{graphicx}
\usepackage{dcolumn}
\usepackage{bm}
\usepackage{amsmath}
\usepackage{amsfonts}
\usepackage{amssymb}
\providecommand{\U}[1]{\protect\rule{.1in}{.1in}}
\begin{document}

\preprint{}

\title{Transition from soft- to hard-Pomeron in the 
structure functions of hadrons at small-$x$ from holography}%

\author{Akira Watanabe}%

\email{j1210709@ed.tus.ac.jp}%

\author{Katsuhiko Suzuki}%

\email{katsu\_s@rs.kagu.tus.ac.jp}%

\affiliation{Department of Physics, Tokyo University of Science, Shinjuku, Tokyo 162-8601, Japan}%

\date{\today}%

\begin{abstract}%

We study the nucleon and pion structure functions at small 
Bjorken-$x$ region in the 
framework of holographic QCD with a special emphasis on the roles of AdS 
space wave 
functions.  
Using the BPST kernel for the Pomeron exchange and calculating its coupling to 
target hadrons in the AdS space, we obtain 
 $F_2$ structure functions at the small-$x$.  
Results for the proton $F^p_2$ as well as the pion  $F^\pi_2$ are consistent 
with experimental data  of the deep inelastic scattering and the forward 
electroproduction of a neutron.     
Observed $Q^2$ dependence of the Pomeron intercept is well reproduced from 
soft non-perturbatibve $(Q^2 \sim 0)$ to hard perturbative $(Q^2 \gg 
1\mbox{GeV}^2)$ region.  
We find the interplay between soft and hard Pomerons 
is closely related with behavior of AdS wave functions of hadrons and the 
virtual photon.

\end{abstract}



\maketitle

\section{\label{sec:level1}Introduction}

The structure functions of hadrons $F_2(x,Q^2)$ at the small 
Bjorken-$x$ region 
provide unique opportunities to access 
the gluon dynamics of QCD.  
At $x \le 10^{-2}$, the structure function is dominated by the 
gluon contribution, which may be identified with the Pomeron in 
QCD~\cite{FS}.  
Within the Regge theory, the elastic (or diffractive) forward 
scattering amplitude is 
described by the exchange of the vacuum quantum number, which is so 
called Pomeron.  
Assuming the Pomeron exchange with $\alpha_0$ and $\alpha'$ being 
constants, one can write a two-body scattering 
amplitude
as
\begin{align}
{\cal A}(s,t) \sim s^{\alpha_0 + \alpha't} \; ,
\end{align}
where the energy $\sqrt{s}$ and the momentum transfer $t$ 
satisfy the Regge kinematics $s \gg t$.  
Thus, the total cross section is expressed by 
\begin{align}
\sigma (s) \sim s^{\alpha_0-1} \, ,
\end{align}
which indicates the cross section of the high energy 
scattering depends on a single parameter  
$\alpha_0$, Pomeron intercept.

Phenomenologically, hadron-hadron collision as well as photon-hadron 
scattering at high energies 
are described  very well by the `soft' 
Pomeron intercept 
$\alpha_0 \sim 1.1$~\cite{DL}.  
However, some hard scale $Q$ enters into the process, situation 
becomes different.  
For example, in the deep 
inelastic scattering (DIS) off the nucleon with the 
photon virtuality $Q^2$, the Pomeron 
exchange model leads to a form of  the structure 
function $F_2$ as
\begin{align}
F_2 (x,Q^2) \sim x^{1-\alpha_0} \; ,
\label{F2}
\end{align}
which is valid for  
the small Bjorken-$x$, $x=Q^2/s \ll 1$.  
In this hard process, the 
experimental data~\cite{Breitweg:1998dz} is consistent 
with the `hard' intercept $\alpha_0 \sim 1.4$ for $Q^2 \gg  1\mbox{GeV}^2$ in contrast to the soft 
Pomeron value $\alpha_0 \sim 1.1$.  
This large value 
of the intercept 
is rather  consistent with the BFKL Pomeron, which is the Reggeized two gluon exchange calculated by 
perturbative QCD~\cite{BFKL}.   
Hence, it is reasonable to conclude that the Pomeron intercept depends 
 on the scale of the scattering process, 
$\alpha_0 (Q)$, where the scale $Q$ means,    
$e.g.$~photon virtuality, quark mass, or momentum transfer.   
Indeed, similar scale dependence is observed in the diffractive photoproduction of light and heavy vector mesons~\cite{Ivanov:2004ax}.

Although we have a clear signal of the transition between `soft' 
and  `hard' Pomerons, 
we cannot calculate such a scale dependence theoretically,  
because it is extremely hard to obtain the soft Pomeron as a 
solution of the non-perturbative QCD.  
Recently, the holographic description of QCD has 
gathered theoretical interests 
as a tool to calculate the non-perturbative quantities in QCD.  
Using the AdS/CFT correspondence (or gauge/gravity in general), 
one may relates the strong coupling 
gauge theory at the boundary with the classical theory of the gravitation 
in the bulk AdS 
space in the large 't~Hooft coupling limit~\cite{Maldacena:1997re,Gubser:1998bc,Witten:1998qj}.  
According to the AdS/CFT dictionary,  
the holographic description of QCD may be achieved by the top 
down approach which originates from the 
string theory in the higher dimensional space~\cite{review}, or more 
phenomenological bottom up approach~\cite{SonStephanov,Erlich:2005qh}.   
The holographic models are applied to 
phenomenological studies successfully~\cite{review,{Polchinski:2001tt},review_meson,review_bottom}.

The first study of the Pomeron with the gauge/gravity 
correspondence was 
done in Ref.~\cite{Brower:2006ea}.  Later, 
much elaborated 
studies have been done in Refs.~\cite{Brower:2007qh,
Brower:2007xg,Brower:2010wf}.  
In those studies, the 
Brower-Polchinski-Strassler-Tan (BPST) kernel which represents the Pomeron exchange 
in the AdS space is introduced based on the string theory 
to describe $Q^2$ dependence of the Pomeron intercept.  
In Ref.~\cite{Brower:2010wf}, the authors calculated the nucleon structure function at the 
small-$x$ 
with the `super local approximation', 
in which overlap functions (probability distributions in the AdS space) of the photon and the nucleon are simply replaced with delta functions.
The 
results are in good agreement with the data, although the validity of 
this approximation seems to be unclear.

In this paper, we shall calculate the nucleon structure function 
at the small-$x$ based on the BPST Pomeron kernel without resorting 
the `super local approximation'.   
To do so we use the AdS wave functions of hadrons calculated 
from holographic QCD, 
and obtain their coupling to the Pomeron in the AdS space.    
Methods to deal with the nucleon properties 
have been already developed and can be applied to the 
Pomeron exchange.  
We demonstrate a 'consistent' holographic description of the structure 
function, by calculating the Pomeron kernel, wave functions 
of scattering particles, and the Pomeron-hadron-hadron three point function, 
gives excellent results for the small-$x$ structure function  
without any artificial assumptions.

In addition, we calculate the pion structure function at 
the small-$x$ in the same way.
We find $F^\pi_2(x)$ at the small-$x$ is suppressed compared with the nucleon case.  
In spite of  the lack of the DIS data for the pion,  
one can extract (model dependent) 
information on the pion structure function from the 
forward electro-production of a neutron $e+p\to e'+n+X$ with a 
large rapidity gap~\cite{Aaron:2010ab}.  
Our calculations are in good agreement with the existing data.

This paper is organized as follows.
In Sec.~\ref{sec:level2}, we introduce the BPST kernel and discuss its properties with the 
delta function approximation for the overlap functions.    
We show why behavior of the AdS wave functions is important to reproduce the scale 
dependence of the effective Pomeron intercept.  
Sec.~\ref{sec:level3} is devoted to the short review of calculating 
the wave functions of hadrons and their couplings with the Pomeron in 
the holographic QCD.   
We show numerical results for the proton structure function in 
Sec.~\ref{sec:level4}.  
In Sec.~\ref{sec:level5}, we calculate the pion structure function and compare 
it with 
the existing parameterizations 
and the data of the forward neutron production at 
HERA.  
Final section is devoted to the summary and discussions.

\section{\label{sec:level2}BPST kernel for Pomeron exchange 
in AdS space}

We first introduce the kernel in the AdS space for the Pomeron exchange, 
and 
discuss its properties in some detail.  
We assume the two-body scattering amplitude ${\cal A}$ for a process 
$1+2 \to 3+4$ at the high energy is given by~\cite{Brower:2006ea,Brower:2007qh,Brower:2007xg} 
\begin{align}
{\cal A} (s,t) = 2is \int d^2b \, e^{i \bm{k_\perp } \cdot \bm{b}} \int dzdz'  P_{13}(z) P_{24}(z') \{ 1-e^{i \chi (s,b,z,z')} \}\;\; ,
\end{align}
where $s$ is the invariant mass square $s=(p_1+p_2)^2$, and $t$ 
the momentum transfer $t=(p_1-p_2)^2$.  
This expression is valid for  the near-forward scattering 
with a condition $s \gg t$, and is dominated by the 
Pomeron exchange.  
The impact parameter $\bm{b}$ is the transverse vector perpendicular to the forward direction.   
The BPST Pomeron exchange kernel is expressed as 
the eikonal form $1-e^{i \chi (s,\bm{b},z,z')}$.  
$P_{13}(z) $ and $P_{24}(z')$ are overlap functions of incoming and 
outgoing particles with their 5D-coordinates $z$ and $z'$.   
Physically, the overlap functions stand for the 
density 
distribution functions of participants in the AdS space, as we will see later.

We concentrate on the deep inelastic scattering in this work.  Thus, it 
is enough to consider the virtual photon $\gamma^*$ $(1 = 3 = \gamma^*)$ 
with the virtuality $Q^2$ 
and the nucleon $(2 = 4 = N)$ in the forward limit $t=0$.  
Using the optical theorem and keeping the leading 
contribution in the eikonal approximation for the kernel, we write the 
structure function 
at the small $x = Q^2/s$ as 
\begin{align}
F_2(x,Q^2)= \frac{Q^2}{2 \pi ^2} \int d^2b \int dzdz' P_{13}(z,Q^2) P_{24}(z') \;
\mbox{Im} \chi (s,\bm{b},z,z') \; ,
\end{align}
where the imaginary part of the BPST kernel is given by
\begin{align}
\mbox{Im} \chi (s,\bm{ b},z,z') = \frac{g_0^2}{16\pi} \sqrt{\frac{\rho}{\pi}} e^{(1-\rho)\tau} 
\frac{\xi}{\sinh \xi} \frac{\exp (\frac{-\xi ^2}{\rho \tau})}{\tau^{3/2}} \; , 
\label{kernel_conf}
\end{align}
where
\begin{align}
\tau &=\log (\rho zz' s/2) \;, \\
\xi &= \sinh ^{-1} \left( \frac{b^2+(z-z')^2}{2zz'} \right) \; .
\end{align}
Here, $g_0^2$ and $\rho$ are the parameters of the model, which are specified later.  
Carrying out the integration over $\bm{b}$ analytically, one finds 
\begin{align}
\int d^2b \;  \mbox{Im} \chi (s,\bm{b},z,z') = \frac{g_0^2}{16} \sqrt{\frac{\rho^3}{\pi}} 
(zz') \; e^{(1-\rho)\tau} \frac{\exp (\frac{-(\log z - \log z')^2}{\rho \tau})}{\tau^{1/2}} \; .
\end{align}
Hence, one can write down an expression for $F_2(x,Q^2)$ as
\begin{align}
F_2(x,Q^2) = \frac{g_0^2 \rho^{3/2}}{32 \pi ^{5/2}} \int dzdz' P_{13}(z,Q^2) P_{24}(z') \;  (zz'Q^2) e^{(1-\rho)\tau} \frac{\exp (\frac{-(\log z - \log z')^2}{\rho \tau})}{\tau^{1/2}} \; .
\label{f2conformal}
\end{align}

The expression (\ref{f2conformal}) is obtained within 
the conformal field 
theory.  
However, we need some energy (or length) scale  which breaks the conformal invariance to reproduce realistic QCD 
in the 4-dimension.  
One of the simplest choices to break the conformal symmetry 
is to introduce the sharp cutoff  
for the AdS coordinate $z$.  
The `hard-wall' model with a sharp cutoff $z_0$ provides with 
a confinement and a mass gap of the hadron spectrum for the 
holographic QCD.   
This model has been applied successfully 
to calculations of various 
observables of 
QCD~\cite{Polchinski:2001tt}.

\begin{figure}[bt]
\includegraphics[width=90mm]{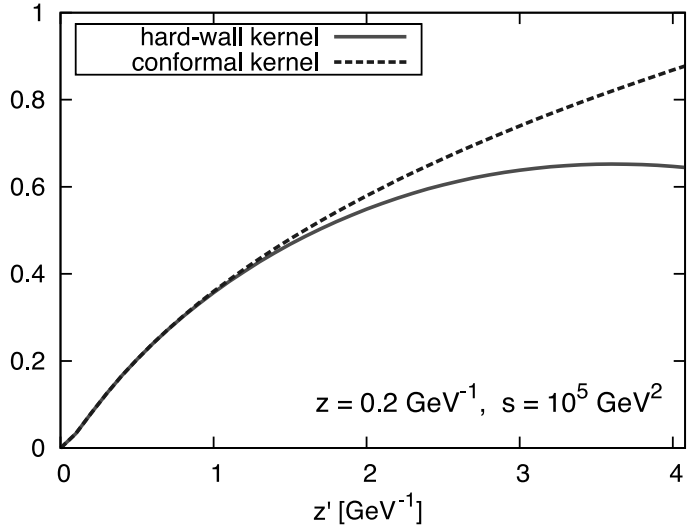}
\caption{ Comparison of BPST kernel.  The solid and dashed curves indicate the hard-wall and conformal kernels, respectively.   Here, we use $\rho = 0.77$.  
}%
\label{fig:kernelz}%
\end{figure}

Inclusion of the hard-wall in the BPST kernel is 
already considered~\cite{Brower:2010wf},   
by the following simple substitution of the Pomeron exchange kernel, 
$\chi_c \to \chi_{\mbox{hw}} $, 
\begin{align}
&\mbox{Im} [\chi_c(s,z,z') ] \equiv 
e^{(1-\rho)\tau} \;e^ {-\frac{\log ^2 z/z'}{\rho \tau}} / {\tau^{1/2}} \; ,
\label{kernel_con1} \\
&\mbox{Im} [\chi_{\mbox{hw}}] \equiv 
\mbox{Im} [\chi_c ] +  \mathcal{F} (z,z',\tau) \, \mbox{Im} [\chi_c(s,z,z_0z_0'/z') ] \; ,\label{kernel_hard} \\
&\mathcal{F} (z,z',\tau ) = 1 - 2 \sqrt{\rho \pi \tau} e^{\eta^2} \mbox{erfc}( \eta ), \nonumber \\
&\eta = \left( -\log \frac{zz'}{(z_0 z_0')} + \rho \tau \right) / {\sqrt{\rho \tau}} \nonumber \; , 
\end{align}
where $z_0$ is the hard-wall parameter.  
With this substitution, $F_2$ now has a form
\begin{align}
F_2(x,Q^2) = \frac{g_0^2 \rho^{3/2}}{32 \pi ^{5/2}} \int dzdz' P_{13}(z,Q^2) P_{24}(z') (zz'Q^2)
\; \mbox{Im} [\chi_{\mbox{hw} }] \;\; .
\label{f2hard}
\end{align}
To study qualitative difference 
between the conformal (\ref{kernel_con1}) and hard-wall kernel 
(\ref{kernel_hard}), we show in Fig.~\ref{fig:kernelz}
$\mbox{Im} [\chi_c ]$ and $\mbox{Im} [\chi_{\mbox{hw}} ]$ as a 
function of $z'$ 
with the $z=0.2 \mbox{GeV}^{-1}$ (typical value of $z$ for 
the virtual photon overlap function $P_{13}$).       
While both Pomeron kernels show the same behavior around small $z'$, 
the hard-wall model deviates from the conformal one  
for $z' \sim 3 \mbox{GeV}^{-1} \sim 0.6 \mbox{fm}$, which may 
correspond to 
a typical size of the hadrons.  
Later, we show numerical results of both hard-wall and conformal 
cases to discuss the role of the hard-wall boundary.

The advantage of the use of the BPST kernel is to incorporate the 
correct scale dependence of the Pomeron intercept~\cite{Brower:2006ea,Brower:2007qh,Brower:2007xg}.  
As we have discussed in the Introduction, the Pomeron intercept 
$\alpha_0$ is close to 1 for the soft hadronic interaction.  
When the hard scale appears in 
the process like DIS, $\alpha_0$ 
increases up to about 1.4. 
Therefore, it is important to discuss 
how the energy scale of the process 
changes  the Pomeron intercept in this model.

In order to see such properties of the Pomeron kernel, it is 
 convenient to adopt `super local approximation' for the
overlap functions~\cite{Brower:2010wf}.  
The `super local' ansatz is nothing but 
the delta function approximation for the distribution functions in the 
AdS space,
\begin{align}
P_{13}(z,Q^2) \approx \delta (z-1/Q) \; , \;\;\;  P_{24}(z') \approx \delta (z'-1/Q') \;  ,
\end{align}
where $Q$ and $Q'$ may be understood as
the typical energy scales of the incident 
and target particles.  
Thus, for DIS, $Q$ and  $Q'$ may be identified with the virtual 
photon momentum and the mass of the target hadron, respectively.

\begin{figure}[bt]
\includegraphics[width=90mm]{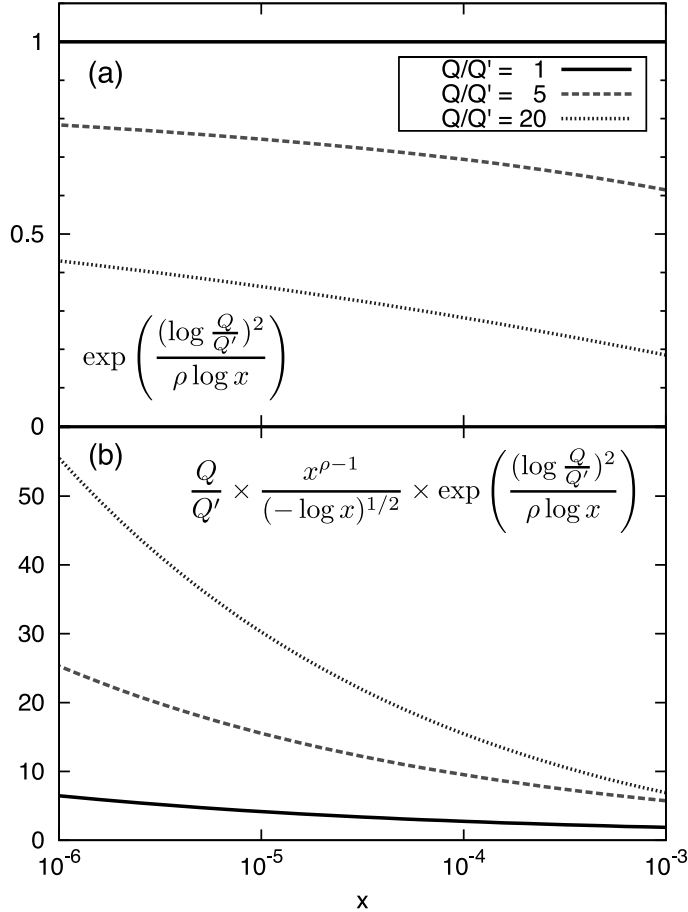}
\caption{ $F_2$ with the delta function approximation for the wave functions.  
(a) Last exponential factor in Eq.~(\ref{kernelx}) as a function of $x$ for $Q/Q'$ .  
(b) $F_2$ for $Q/Q'=1,5,20$.  
}%
\label{fig:kernelx}%
\end{figure}

Inserting them into (\ref{f2conformal}) one obtains (here we consider 
only the conformal case for simplicity.)
\begin{align}
F_2(x,Q^2) \propto \frac{Q}{Q'} \times 
\frac{x^{\rho-1}}{ ( -\log x ) ^{1/2} }\times 
\exp \left( \frac{(\log \frac{Q}{Q'} )^2}{\rho \log x } \right)\; .
\label{kernelx}
\end{align}
In Eq.~(\ref{kernelx})  the Bjorken-$x$ dependence 
mainly comes from a term 
$x^{\rho-1} / ( -\log x ) ^{1/2}$, which is independent of $Q,Q'$. 
The subtle $Q,Q'$ dependences arise from the 
last exponential factor 
$\exp \left( \log ^2 \frac{Q}{Q'} / \rho \log x \right)$.
Fig.~\ref{fig:kernelx} shows the $x$-dependences of Eq.~(\ref{kernelx}) for 
several values of $Q,Q'$.  
With $Q/Q' = 1$ it shows rather weak $x$ dependence, where the resulting Pomeron 
intercept is small, $\alpha _0 \sim 1.1$.  
If we set 
 $Q/Q '= 20$, the BPST kernel increases 
rapidly as $x$ decreasing, giving the large intercept, $\alpha_0 \sim 1.4$.  
This striking change is due to $\log(Q/Q')$ factor in the 
exponential factor shown in Fig.~\ref{fig:kernelx}.  
Hence, in order to require that the BPST kernel 
is capable of describing the `soft' to `hard' transition of the Pomeron, 
$Q/Q' $ must be large for the hard process, while $Q/Q' \sim 1$ for 
the soft process.

In the super local approximation, 
$1/Q,1/Q'$ are simply related with   
positions of the delta function peak for  
$P_{13}(z)$ and $P_{24}(z')$.  
For example, if we want to acquire the large Pomeron intercept, 
we need a 'gap' between the peak positions of the 
overlap functions of the virtual photon and the nucleon, namely, $P_{13}(z)$ 
must have a sharp peak at $z \sim 0$, which is far away from the distribution of 
$P_{24}(z')$ in the AdS space.  
In the next section we introduce the AdS wave functions for the nucleon and 
pion, and 
calculate their overlap functions.    
We show such a `gap' in the AdS space  between $P_{13}$ and $P_{24}$  
naturally incorporated
within the holographic QCD as the scale $Q^2$ increases.

\section{\label{sec:level3}Hadronic wave functions in  AdS space and 
Pomeron coupling}

In order to calculate Eqs.~(\ref{f2conformal}) and (\ref{f2hard}) 
we need evaluate the hadronic wave functions in the AdS 
space in terms of the holographic QCD.   
For the virtual photon, we use the massless 5D U(1) vector field~\cite{Polchinski:2002jw,Brower:2010wf,Hatta:2007he,Levin:2009vj},  
which is dual to the electromagnetic current.  
This vector fields satisfy the Maxwell equation in the bulk AdS space.  
The overlap function of the virtual photon is found to be~\cite{Brower:2010wf}
\begin{align}
P_{13}(z,Q^2) = \frac{1}{z} (Qz)^2 \left( K_0^2(Qz) + K_1^2(Qz) \right) \;\; .
\end{align}
This function in the AdS space basically corresponds to the photon 
impact factor in  QCD.

On the other hand, we adopt the hadronic wave functions calculated in the 
AdS space
with the metric, 
\begin{align}
ds^2  = g_{MN} dx^M dx^N  = \frac{1}{{z^2 }}\left( {\eta _{\mu \nu } dx^\mu  dx^\nu   - dz^2 } \right) \; . 
\end{align}
We consider the hard-wall model $\varepsilon \leq z \leq z_0$ $(\varepsilon \to 0)$ in  the following 
calculation where IR boundary $z_0$ breaks the 
conformal symmetry to mimic the realistic QCD.  
Hereafter, we briefly review the calculations of the AdS wave functions in the 
holographic QCD.

For the nucleon case, we use the model of 
Refs.~\cite{Henningson:1998cd,Muck:1998iz,Abidin:2009hr,Contino:2004vy,Hong:2006ta} in which the 
nucleon is described 
by a solution of the 5-dimensional Dirac equation.  
The classical action for the Dirac field is given by
\begin{align}
S_{F}  = \int {d^5 x} \sqrt{g} \biggl( \frac{i}{2} \bar{\Psi} e_A^N \Gamma^A D_N \Psi 
- \frac{i}{2} (D_N \Psi )^\dag \Gamma^0 e_A^N \Gamma^A \Psi - M  \bar{\Psi} \Psi \biggr) \; ,
\end{align}
where $e^N_A=z \delta^N_A$ and $D_N=\partial_N 
+\frac{1}{8}\omega_{NAB}[\Gamma^A,\Gamma^B]-iV_N$
are introduced to maintain the invariance of the action under the gauge 
transformation and 
generalized coordinate transformation in the AdS space.   
The mass of the spinor is $M=3/2$ by the AdS/CFT dictionary.  
${\Psi}$ is obtained as a solution of the Dirac equation,%
\begin{equation}
\left( i e^N_A \Gamma^A D_N - M \right) \Psi = 0 \;\;.
\end{equation}
To derive this equation we require a boundary condition at the hard-wall, either $\Psi_R(z_0) = 0$ or  $\Psi_L (z_0) = 0$, where 
we define the right- and left-handed spinor; 
$\Psi_{R,L}=(1/2)(1\pm \gamma^5)\Psi$.

Carrying out the Fourier transform in the $4$-dimensional space with a 
momentum $p^\mu$, one can express the Dirac field as 
$$
\Psi_{R,L} (p,z) = z^\Delta \, \Psi^0_{R,L} (p) f_{R,L} (p,z) \;\; ,   
$$
where $\Psi^0_{R,L} (p) $ is a plane wave solution in the 4-dimensional space which plays a role of the source 
field for the spin-1/2 baryon operator, and 
$f_{R,L} (p,z) $ the bulk-to-boundary propagator.  
$\Delta$ is chosen to satisfy a condition $f(p,\varepsilon) = 1$.

Dropping the interaction term with the gauge field, one can rewrite the 
Dirac equation as, 
\begin{align}
\left(\partial_z - \frac{2+M-\Delta}{z}\right)f_R& =-pf_L \; , \nonumber \\
\left(\partial_z - \frac{2-M-\Delta}{z}\right)f_L& =pf_R \; ,
\label{eosN}
\end{align}
where $p=\sqrt{p^2}$.  
Imposing a condition that solutions are not singular at $z=\varepsilon$, 
we require $\Delta = 2 - M$.  

Choosing the 
boundary conditions $f_R (z_0) = 0$ and $f_L(\varepsilon) = 1$, 
one finds normalizable modes $\psi_L (z) = f_L(p=m_n^N,z)$ and $\psi_R 
(z)  = 
f_R (p=m_n^N,z)$ from Eqs.~(\ref{eosN}) with $m^N_n$ being 
the mass of the $n$-th Kaluza-Klein state; 
\begin{align}
\psi_L^{(n)} (z) = \frac{\sqrt{2} z^\alpha J_\alpha (m_n^N z)}{z_0 J_\alpha (m_n^N z_0)} \; , \; \; \; 
\psi_R^{(n)} (z) = \frac{\sqrt{2} z^\alpha J_{\alpha - 1} (m_n^N z)}{z_0 J_\alpha (m_n^N z_0)} \;  .
\end{align}
Here, $\alpha = M+\frac{1}{2}$ and the eigenvalue $m_n^N$ is determined by the boundary 
condition at IR boundary,  $J_{\alpha-1}(m_n^N z_0)=0$.  
The hard-wall parameter $z_0$ is fixed to reproduce the nucleon mass 
as the lowest eigenstate.  
These solutions are subject to the normalization conditions,
\begin{align}
\int dz\, \frac{1}{z^{3}} \psi^{(n)}_{L}\psi^{(m)}_{L} = 
\int dz\, \frac{1}{z^{3}} \psi^{(n)}_{R}\psi^{(m)}_{R} = \delta_{nm} \; .
\label{nucleon-norm}
\end{align}

On the other hand, we adopt the pion wave function studied in 
Ref.~\cite{Erlich:2005qh}, which will be sketched below.   
This wave function is successfully applied to the calculation of 
dynamical quantities of the pion, {\em e.g.}~electromagnetic form 
factor~\cite{Kwee:2007dd,Grigoryan:2007wn} or anomalous $\pi^0 \to \gamma \gamma$ transition~\cite{Grigoryan:2008up}.   
The effective action for the meson fields is 
\begin{align}
S_{M}  = \int {d^5 x} \sqrt g \biggl[ \mathrm{Tr} \biggl\{ \left| {DX} \right|^2  + 
3\left| X \right|^2  - \frac{1}{{4g_5^2 }}\left( {F_L^2  + F_R^2 } \right) \biggr\} \biggr] \; ,
\end{align}
where the covariant derivative $D^M X=\partial^M X-i A_L^M X+iX A_R^M$
is introduced with the gauge field $A_L^M, A_R^M$, and 
$F^{MN}_{L,R}=\partial^M A^N_{L,R}-\partial^N A^M_{L,R}-i[A^M_{L,R},A^N_{L,R}]$, 
the field strength tensor.  
The bulk field $X$ is defined by $X(x,z)=X_0(z) \exp (2i t^a \pi^a)$
with the bulk scalar $X_0$ and the pion field $\pi^a$ with the 
isospin operator $t^a$.  
The bulk scalar is given by
\begin{align}
X_0=\frac{1}{2}\mathbb{I}v(z) =\frac{1}{2}\mathbb{I} \left( m_q z+\sigma z^3 \right) \; ,
\end{align}
where $m_q$ can be identified with the quark mass and $\sigma$ the chiral condensate.  
Hereafter, we take the chiral limit, $m_q = 0$, for simplicity, and   
the pion is always massless.

Up to the second order of the fields, the action for the pion $\pi$ 
and the axial-vector field $A=(A_L-A_R)/2$ is 
\begin{align}
S_A = \int d^5 x \sqrt{g} \biggl[ \frac{v(z)^2}{2} g^{MN} (\partial_M\pi^a-A^a_M) (\partial_N\pi^a-A^a_N) - \frac{1}{4g_5^2} g^{KL} g^{MN}  F^a_{KM}F^a_{LN}   \biggr] \; .
\label{action-pi}
\end{align}
We work with the gauge fixing condition $A_z=0$ for the axial-vector field.  
The axial vector field can be  expressed as 
$A^a_\mu  ={A^{a}_\mu}_\perp  + {A^{a}_\mu}_\parallel$, where 
${A^{a}_\mu}_\perp$ and ${A^{a}_\mu}_\parallel$ are the transverse and 
longitudinal components.   The longitudinal part 
${A^{a}_\mu}_\parallel = \partial_\mu \phi^a $ contributes to the Goldstone mode due 
to the spontaneous chiral symmetry breaking.

Applying the Fourier transform with a momentum $q$, 
we decompose $A_\perp, 
\phi, \pi$ fields into sources in the 4-dimension 
and the bulk-to-boundary propagators $A(q,z)$, $\phi(q,z)$, and $\pi(q,z)$.  
One obtains classical equations of motion for the  pion sector as 
\begin{align}
&\partial_z \left(\frac{1}{z}\partial_z {A(q,z)} 
\right)+\frac{q^2}{z}{A(q,z)} -\frac{g_5^2v^2}{z^3}{A(q,z)} =0 \; ,
\label{eom_a1}\\
&\partial_z \left(\frac{1}{z}\partial_z \phi(q,z) \right) +\frac{g_5^2v^2}{z^3}\left(\pi (q,z)-\phi (q,z)\right)=0 \; , 
\label{eom_pion_1}\\
&-q^2\partial_z \phi (q,z)+ \frac{g_5^2v^2}{z^2}\partial_z \pi(q,z)=0 
\; . 
\label{eom_pion_2}
\end{align}
We look for solutions of Eqs.~(\ref{eom_a1},\ref{eom_pion_1},\ref{eom_pion_2}), which 
satisfy the  boundary conditions, $\phi(0)=0$, 
$\partial_z \phi(z_0)=0$, and  $\pi(0)=0$~\cite{Erlich:2005qh}.  
In addition, $\pi(z)$ must approach $-1$ away from $z=\varepsilon$ to 
recover the Gell-Mann$-$Oakes$-$Renner relation~\cite{Erlich:2005qh}.  
On the other hand, the bulk-to-boundary propagator of the transverse axial-vector field 
$A(q,z)$ is subject to $A(q,\varepsilon) = 1$ and $\partial _z A(q,z_0) =0$.

Since we consider the massless pion in the chiral limit, $m_\pi^2 = q^2 = 0$, 
we obtain $\pi(0,z) = -1$ from 
Eq.~(\ref{eom_pion_2}) and the boundary condition.  
In this case, Eq.~(\ref{eom_pion_1})   coincides with 
Eq.~(\ref{eom_a1}), giving a relation $\phi(0,z) = A(0,z) -1 =A(0,z) + \pi(0,z)
 $.  
Thus, it is convenient to introduce
 the new wave function of the pion, $\Psi(z) \equiv \phi(z)-\pi (z) = A(z)$~\cite{Grigoryan:2007wn}.  
We write the equation for $\Psi$ from Eq.~(\ref{eom_pion_1}), 
\begin{equation}
z^3 \partial _z \left( {\frac{1}{z}\partial _z \Psi } \right) - g_5^2 v^2 \Psi  = 0 \; ,
\label{pion_mod}
\end{equation}
which leads to a solution, 
\begin{align}
\Psi \left( z \right) = z\Gamma \left[ {\frac{2}{3}} \right] \left( {\frac{\beta }{2}} \right)^{1/3} \biggl[ I_{ - 1/3} \left( {\beta z^3 } \right) - I_{1/3} \left( {\beta z^3 } \right)\frac{{I_{2/3} \left( {\beta (z_0^\pi)^3 } \right)}}{{I_{ - 2/3} \left( {\beta (z_0^\pi)^3 } \right)}} \biggr] \; ,
\end{align}
with  $\beta= 2\pi \sigma/3$.

On the other hand, one can relate $A(0,z)$ with the pion decay constant $f_\pi$, 
defined by $\langle 0 | A_\mu | \pi(q) \rangle = i f_\pi q_\mu$.  
Calculation of the axial-current correlator  based on the holographic 
principle yields,
\begin{align}
 f_\pi^2 = - \left. 
\frac{1}{g_5^2} \frac{\partial _z A(0,z) }{z} \right|_\varepsilon \;\; .
\label{piondecay}
\end{align}

The model parameters $\sigma$ and  IR hard-wall $z_0$ can be 
determined to reproduce the 
$\rho$ meson mass  and the pion decay constant~\cite{Erlich:2005qh}.  
The gauge coupling $g_5$ is fixed by the matching condition at the 
UV boundary $z=\varepsilon$.

In order to calculate the Pomeron coupling with the hadrons, we consider 
the hadron-graviton-hadron three point function
~\cite{Abidin:2008hn,Abidin:2009hr}.  
To do so, we introduce the perturbation to the metric 
$\eta_{\mu\nu}\to\eta_{\mu\nu}+h_{\mu\nu}$ in the classical action, 
and extract $h AA$ terms, which contribute to the three point function.

First we determine behavior of $h_{\mu\nu}$ itself by solving the 
equation 
of motion.  
Introducing the 4-momentum $q$ for $h_{\mu\nu}$ after the Fourier 
transform, we 
write $h_{\mu\nu}(q,z) =h^0 _{\mu\nu}(q) H (Q,z)$ with 
$Q=\sqrt{-q^2}$.  
By virtue of the 
transverse-traceless gauge, $\partial^\mu h_{\mu\nu}=0$ and 
$h^\mu_\mu=0$,
the linearized Einstein equation for the bulk-to-boundary propagator 
${H}(Q,z)$ is   
\begin{align}
\left[ \partial_z\left(\frac{1}{z^3}\partial_z\right) +\frac{1}{z^3}q^2\right] {H}(Q,z) =0 \; .
\end{align}
Imposing the conditions, $\partial_z {H}(Q,z_0)=0$ and ${H}(Q,\varepsilon)=1$, 
we obtain
\begin{align}
{H} (Q,z) = \frac{1}{2} Q^2 z^2 \left[ \frac{K_1  (Qz_0) }{I_1 (Qz_0) } I_2 (Qz) + K_2 (Qz) \right] \; ,
\end{align}
which satisfies ${H} (\varepsilon,z) =1$ as expected~\cite{Abidin:2008hn,Abidin:2009hr}.

As an example, we will review the pion case following the work of 
Ref.~\cite{Abidin:2008hn}.  
The three point function of the stress tensor with the 
axial currents can be calculated as, 
\begin{equation}
\langle 0 | {\mathcal T} {J}_5^{a\alpha}(x) \hat{T}_{\mu\nu}(y){J}_5^{b\beta}(w) | 0 \rangle =  \frac{ - 2 \delta^3 S}{\delta A^{a0}_\alpha (x) \delta h^{\mu\nu 0}(y) \delta A^{b0}_\beta(w)}  .
\label{pi-three}
\end{equation}
To calculate the RHS of Eq.~(\ref{pi-three}) from the classical action Eq.~(\ref{action-pi}), 
we extract terms which are linear in $h_{\mu \nu}$ and quadratic in 
$\pi$ and $A$;
\begin{align}
S_A^{(3)} = \int {d^5 x} \biggl[ - \frac{v(z)^2 h^{\rho\sigma} }{2z^3} (\partial_\rho\pi^a-A^a_\rho)(\partial_\sigma\pi^a-A^a_\sigma) + \frac{1}{2g_5^2z} h^{\rho\sigma} [-F_{\sigma z}F_{\rho z}+\eta^{\alpha\beta}F_{\sigma\alpha}F_{\rho\beta} ] \biggr] \;.  
\label{action-3}
\end{align}

On the other hand, the LHS of Eq.~(\ref{pi-three}) in the momentum space 
is rewritten 
with the matrix element of the stress tensor 
between the pion states $\langle \pi ^a (p_2) | \hat{T}^{\mu \nu}  | \pi ^b (p_1) \rangle$, 
by inserting the intermediate states (pion) and using 
$\langle 0 | A_\mu | \pi(p) \rangle = i f_\pi p_\mu$ 
with $p_1^2, p_2^2 \to 0$~\cite{Abidin:2008hn}.   
From the symmetry consideration, the matrix element of the stress tensor between the pion states can be parameterized by%
\begin{align}
\langle \pi ^a (p_2) | \hat{T}^{\mu \nu}  | \pi ^b (p_1) \rangle = 2 \delta ^{ab} F_\pi (\ell^2 ) \left [ (p_1+p_2)^\mu (p_1+p_2)^\nu + \cdots  \right ] \; ,
\label{def-gra}
\end{align}
where $\ell = p_2 - p_1$, and  $F_\pi$ determines the 
magnitude of the matrix element.  
Here, we only retain the 
dominant contribution in the small-$x$ kinematics, $p_1,p_2 \gg \ell$.

Functional differentiation of Eq.~(\ref{pi-three}) with 
Eqs.~(\ref{action-3},\ref{def-gra}) gives
\begin{equation}
F_\pi  \left( {\ell^2 } \right) = \int_\varepsilon ^ {z_0}  {dz} 
{H} \left( {\ell^2,z} \right) \left[ \frac{(\partial _z \Psi (z))^2 }{g_5^2 f_\pi ^2 z}   + \frac{v(z)^2 \Psi (z)^2 }{ f_\pi ^2 z^3 }  \right] \; .
\end{equation}
In the case of DIS, it is 
enough to consider the forward limit $\ell=0$, at which 
${H}(\varepsilon,z) = 1$.  
Hence, the overlap function $P_{24}$ for the pion is identified with 
\begin{equation}
P_{24}^\pi (z) = \left[ \frac{(\partial _{z} \Psi (z))^2 }{g_5^2 f_\pi ^2 z}   + \frac{v(z)^2 \Psi (z)^2 }{ f_\pi ^2 z^3 }  \right] \; .
\label{pion-overlap}
\end{equation}

We note that the overlap function Eq.~(\ref{pion-overlap}) satisfies the 
normalization condition; 
$\int ^{z_0}_\varepsilon dz P_{24}^\pi(z)$ $= 1 $.  
One can checnk it by integrating by parts,  
\begin{align}
&\int_\varepsilon^{z_0} dz \left[
\frac{(\partial _{z} \Psi (z))^2 }{g_5^2 f_\pi ^2 z}   + \frac{v(z)^2 \Psi (z)^2 }{ f_\pi ^2 z^3 }  \right] \\ \nonumber
 = & \left[ \Psi (z) \frac{ \partial_{z} \Psi (z)}{g_5^2 f_\pi^2 z} \right]
 ^{z{_0}}_\varepsilon - \int_\varepsilon^{z_0} dz \left[
\frac{\Psi(z)}{g_5^2 f_\pi^2 z^3}  \left\{  z^3 \partial _z \left( 
\frac{\partial _{z} \Psi (z) }{ z} \right) 
  - g_5^2 v^2 \Psi (z)  \right\} \right]  \\ \nonumber
=&1  \; ,
\end{align}
where we use Eqs.~(\ref{pion_mod}) and (\ref{piondecay}), and the 
boundary conditions $\partial _{z} \Psi(z_0) = 0$. 
We also note that 
$P_{24}(z)$ can be understood as the probability density 
distribution in the bulk coordinate 
space, although the expression of Eq.~(\ref{pion-overlap}) involves derivative of the wave function and differs from the familiar $\Psi^2$ form.  However, the 
expression  Eq.~(\ref{pion-overlap}) also appears in the calculations of 
the pion electromagnetic form factor~\cite{Kwee:2007dd,Grigoryan:2007wn} and the $\pi\pi\rho$ coupling constant~\cite{Erlich:2005qh}.  
In fact, 
the expression of Eq.~(\ref{pion-overlap}) depends on the choice of the gauge.  
Alternative gauge fixing condition provides a familiar wave function 
square form~\cite{Grigoryan:2008cc}.

One can apply the similar method to the nucleon case~\cite{Abidin:2009hr}, 
and find 
$P_{24}^N$ as 
\begin{equation}
P_{24}^N (z) = \frac{1}{2z^{3}}  \left( \psi_L^2 (z) + \psi_R^2 (z) \right) \; .
\end{equation}
This is apparently interpreted as the density distribution and 
fulfills the normalization condition~\cite{Abidin:2009hr},  
\begin{equation}
\int_\varepsilon^{z_0} dz P_{24}^N (z) = 1 \; ,
\end{equation}
with the help of the wave function normalization Eq.~(\ref{nucleon-norm}).

\begin{figure}[bt]
\includegraphics[width=95mm]{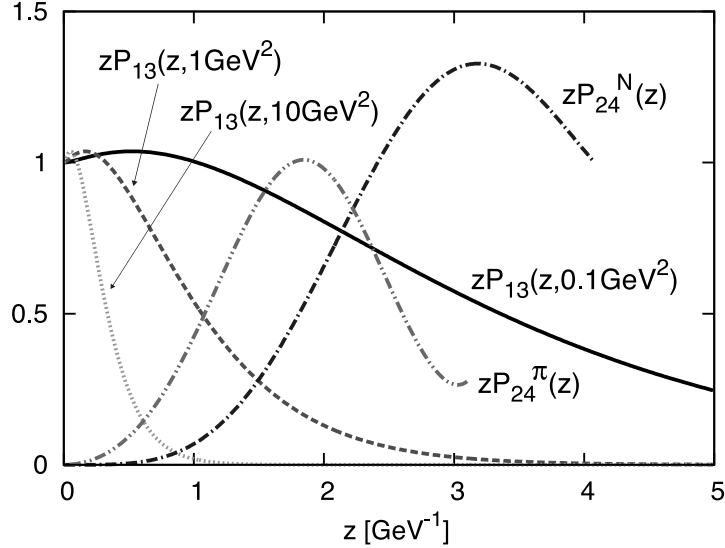}
\caption{The overlap functions in the integrand of Eqs.~(\ref{f2conformal}) and (\ref{f2hard}).  
The solid, dashed, and dotted curves indicate $ zP_{13}(z,Q^2)$ with 
$Q^2 = 0.1 \mbox{GeV}^2$, $1 \mbox{GeV}^2$ and  $10 \mbox{GeV}^2$, respectively.  
$P_{24}(z)$ for the nucleon is shown by the dashed-dotted curve, and one for the pion by the
dashed double-dotted curve.  
}%
\label{fig:overlap}
\end{figure}

We show in Fig.~\ref{fig:overlap}  
the hadronic 
overlap functions $z P_{24}^N (z)$ and $z P_{24}^\pi (z)$, as appeared 
in the formulae of the Pomeron scattering amplitudes, Eqs.~(\ref{f2conformal}) and (\ref{f2hard}).  
Both the nucleon and pion cases  show 
similar $z$-dependences, whose peak positions are located near the hard-wall cutoff 
$z_0$.

In Fig.~\ref{fig:overlap}, 
the overlap functions of 
the virtual photon $P_{13}(z,Q^2)$ with $Q^2 =0.1, 1,$ and $10 
 \mbox{GeV}^2$ are also shown to compare with the nucleon and the 
pion counterparts.  
It is important to note that  
shape of the photon overlap function drastically changes 
as the photon virtuality increases.

In Sec.~\ref{sec:level2} we have shown 
the Pomeron intercept $\alpha_0$ becomes large (hard value) only if  
the distribution of the  photon overlap function, 
$P_{13}(z)$, substantially differs from that of the target, 
$P_{24}(z')$.  
Here, using the realistic AdS wave functions, we find such a 
condition is in fact 
realized only at the larger $Q^2 \sim 10 \mbox{GeV}^2$ region, where $P_{13}$ 
shows a sharp peak at $z\sim 0$.  
On the other hand, Fig.~\ref{fig:overlap} tells us that the photon overlap 
function at lower $Q^2 \leq 1 \mbox{GeV}^2$ extends over the entire 
$z$ region.   Thus, the resulting Pomeron intercept may become smaller   
according to the discussion in 
Sec.~\ref{sec:level2}.  
In the next section we shall demonstrate how the Pomeron intercept $\alpha_0(Q^2)$ changes 
as the scale $Q^2$ increases by the 
numerical calculation.

\section{\label{sec:level4}Numerical Results for Nucleon $F_2^p$}

We first show the numerical results of $F^p_2(x,Q^2)$ at the small-$x$.  
To carry out the numerical calculations, it is necessary to determine the 
parameters of the model, $\rho$ and $g_0^2$.  
The parameter $\rho$ plays a role to fix the energy dependence of the 
cross section.   
We find $\rho = 0.845$ for the conformal case, $\rho = 0.799$ for the hard-wall case.  
The parameter 
$g_0^2$ is determined by the magnitude of the nucleon structure function.  
We find  $g_0^2 = 1.23 \times 10^2$  and $g_0^2 = 1.25\times 10^2$
for the conformal and hard-wall cases, respectively. 
There exist additional parameters, the infrared sharp cutoff $z_0$, 
in the hard-wall model.  For the pion and the nucleon, they are 
determined to reproduce the pion decay constant and the nucleon mass, 
$z_0^\pi = 1/(322 \mbox{MeV})$ and $z_0^N = 1/(245 \mbox{MeV})$, 
respectively~\cite{Erlich:2005qh,Abidin:2009hr}.  
We also introduce the hard-wall cutoff for the virtual photon in Eq.~(\ref{kernel_hard}).  
This parameter is chosen to be $z_0 = 6 \mbox{GeV}^{-1}$ to 
reproduce the DIS data.

\begin{figure}[bt]
\includegraphics[width=140mm]{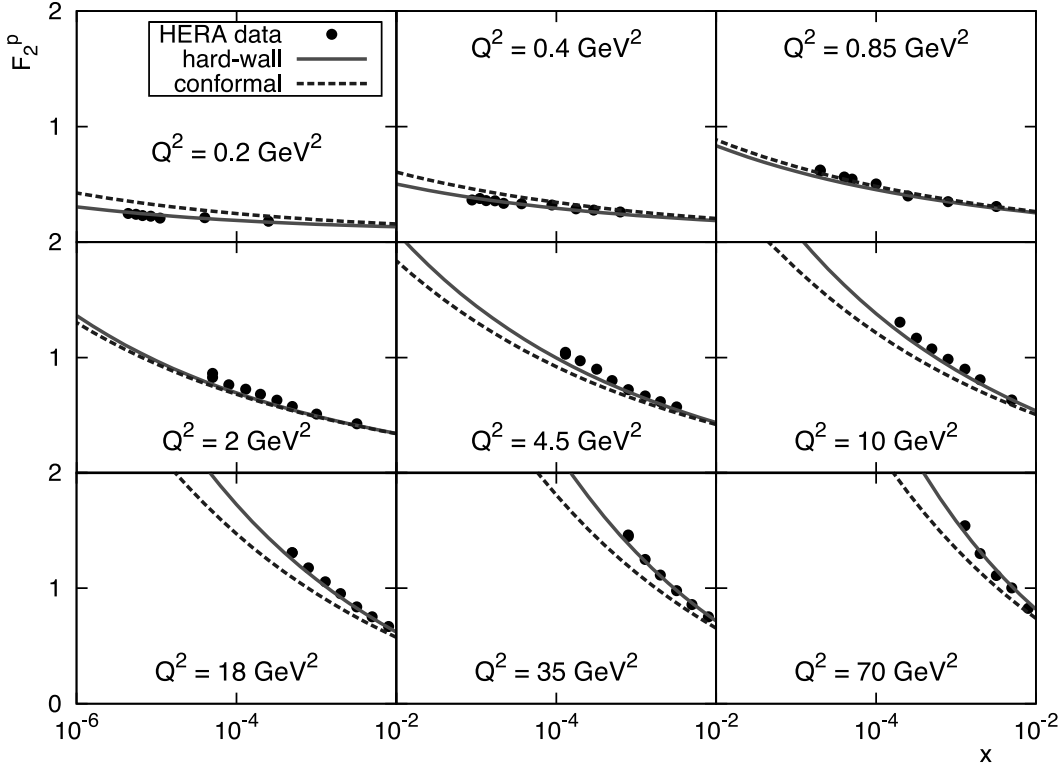}
\caption{ $F^p_2(x,Q^2)$ as a function of the Bjorken-$x$ for various $Q^2$.  In each figure, the solid and dashed curves represent results with hard-wall and conformal kernels, respectively.  HERA data~\cite{Aaron:2009aa} is depicted by circles.
}%
\label{fig:f2pq2}%
\end{figure}

We first show in Fig.~\ref{fig:f2pq2} $F^p_2(x,Q^2)$ as 
a function of $x$ for various $Q^2$. 
At low $Q^2$, $F^p_2$ is rather insensitive to the 
variation of $x$.  However, as $Q^2$ increases, $F^p_2(x)$ rapidly 
grows at the small-$x$ region.  
Calculated results are in good agreement with the HERA data~\cite{Aaron:2009aa}.

If we describe $F_2^p$ at the low-$x$ using the Pomeron exchange with 
the intercept $\alpha_0$, the structure function 
behaves as $F_2^p \sim x^{1-\alpha_0}$.  
From our results in Fig.~\ref{fig:f2pq2} it is easy to recognize  
that  the Pomeron intercept increases as $Q^2$ 
increases,  
which may be understood as a transition from `soft' to `hard' 
Pomerons.

\begin{figure}[bt]
\includegraphics[width=115mm]{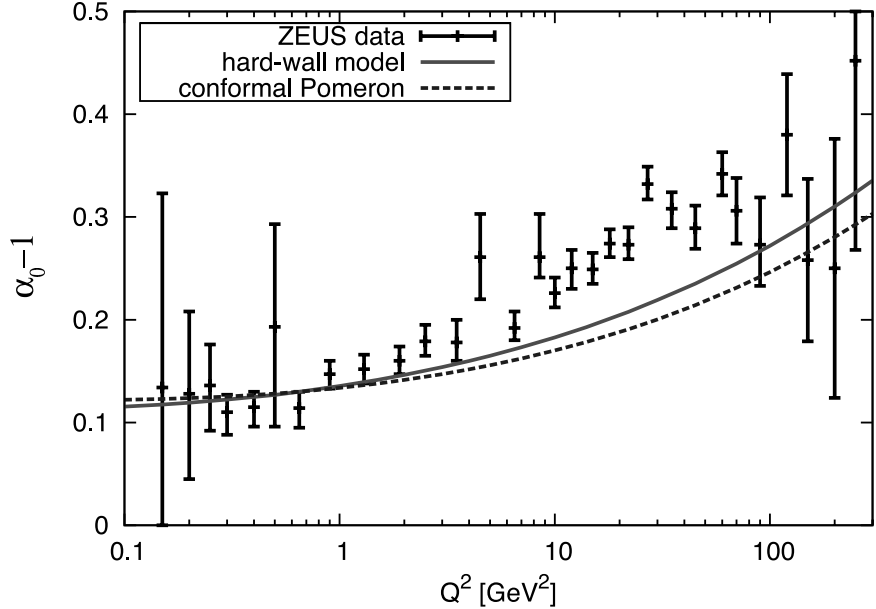}
\caption{ The effective Pomeron intercept $\alpha _0(Q^2)$ for the nucleon. The solid and dashed curves represent results with hard-wall and conformal kernels, respectively.  HERA data~\cite{Breitweg:1998dz} is depicted with error bars.
}%
\label{fig:f2palpha}%
\end{figure}

To see this fact in detail, 
we show in Fig.~\ref{fig:f2palpha} 
 the effective intercept $\alpha_0(Q^2)$ extracted from the 
calculations of $F^p_2$ in Fig.~\ref{fig:f2pq2}.  
It is evident  that 
the grows of $\alpha_0(Q^2)$ with $Q^2$ increasing, although
the magnitude 
is little bit smaller than the data~\cite{Breitweg:1998dz}.  
For the real photon scattering at $Q^2=0$, the calculation shows $\alpha_0 (Q^2)  \sim 1.1$, 
which is consistent with the `soft' Pomeron intercept describing the high energy  
scattering phenomenologically.  On the other hand, $\alpha_0 (Q^2)$ 
reaches about $1.3$ at $Q^2 \sim 100 \mbox{GeV}^2$, 
which is expected from the perturbative `hard' Pomeron picture.

Our results are quantitatively similar with 
those of 
Ref.~\cite{Brower:2010wf} in which the super local approximation is 
adopted.  However, the 
purpose of this work is to clarify whether or not 
the BPST Pomeron kernel with the AdS wave functions calculated from the 
holographic QCD describe the DIS structure function in both soft and 
hard $Q^2$ regions.  
The present results show our framework based on the holography  indeed  works well to 
reproduce the nucleon structure function and account for the scale 
dependence of the Pomeron 
intercept without any artificial assumptions.

Size of the effects introduced by the hard-wall cutoff for the BPST 
Pomeron 
kernel is seen in Fig.~\ref{fig:f2pq2} by the solid curve to be compared
with the conformal 
case (dashed curve).  Whereas the hard-wall results are somewhat smaller 
than those of the conformal kernel at lower $Q^2$, the 
calculation of the hard-wall model 
exceeds the conformal cases  at the higher $Q^2$ region.  
This behavior certainly depends on the 
choice of the parameters, especially $\rho$.  
If we used the same values of $\rho$ and $g_0$ for both 
the hard-wall and  conformal models, 
the solid curve would tend to agree with the 
dashed one at the $Q^2 \to \infty$ limit.  
In any case, the hard-wall model is more 
sensitive to the variation of $Q^2$, and thus adequate to account for 
the 
soft to hard transition of the Pomeron.

\section{\label{sec:level5}Results for Pion Structure Function $F_2^\pi$}

We then apply the 
same framework to the pion structure function $F_2^\pi$.  
The experimental information is not 
enough to understand the 
small-$x$ behavior of the  pion structure function, 
because one cannot perform the deep 
inelastic scattering with the 
pion target.  
The valence quark distribution of the  
pion is  relatively well determined by using the Drell-Yan 
process~\cite{Sutton:1991ay,Gluck:1999xe}.  
However, recent reanalysis with the 
resummation technique indicates there still exist ambiguities
of the shape of the valence distribution~\cite{Aicher:2010cb}.  
Moreover, the sea quark and gluon distributions 
are ill determined.   
In our approach, 
we predict $F^\pi_2$ at the small-$x$ without 
any adjustable parameter, once  
the model parameters, $\rho,g_0$, are
 fixed by the nucleon data.

\begin{figure}[bt]
\includegraphics[width=100mm]{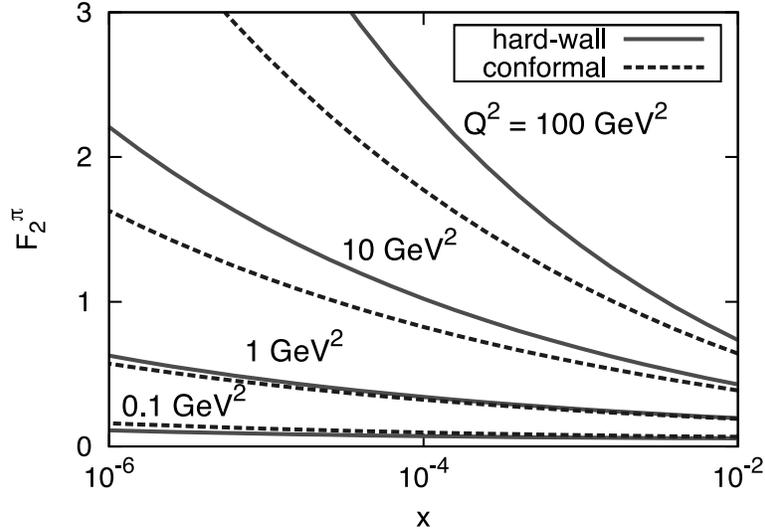}
\caption{ $F^\pi_2(x,Q^2)$ for $Q^2 = 0.1, 1, 10, 100 \mbox{GeV}^2$ are shown as a function of 
$x$.  The solid and dashed curves 
show the hard-wall and conformal results, respectively.  }%
\label{fig:f2pix}%
\end{figure}

\begin{figure}[bt]
\includegraphics[width=100mm]{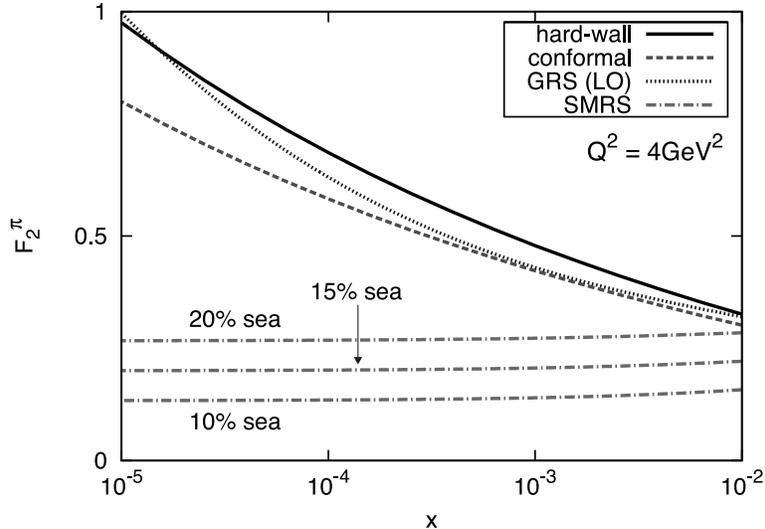}
\caption{ $F^\pi_2(x,Q^2)$ at $Q^2 = 4 \mbox{GeV}^2$.  The solid and dashed curves 
show the conformal and hard-wall results, respectively.  
The SMRS~\cite{Sutton:1991ay}  and GRS~\cite{Gluck:1999xe} parameterizations 
are depicted by the 
dash-dotted and dotted curves, respectively.  }%
\label{fig:f2piSMRS}%
\end{figure}

In Fig.~\ref{fig:f2pix} we show the $x$-dependence of 
$F^\pi_2(x,Q^2)$ for various $Q^2$.  
At $Q^2 = 0.1 \mbox{GeV}^2$, where non-perturbative dynamics dominates, 
the structure function is almost constant.   
With the virtuality $Q^2$ increasing, 
the $x$-dependence becomes steeper,  which is already seen in the 
nucleon case.  
In Fig.~\ref{fig:f2piSMRS} 
the calculations at $Q^2 = 4 \mbox{GeV}^2$ are compared with SMRS~\cite{Sutton:1991ay} and 
GRS~\cite{Gluck:1999xe} parameterizations of 
the pion structure function, and   
shown to be significantly  different from them.
However, 
we could not draw strong conclusion from this comparison, since 
the parameterization of $F^\pi_2$ at the small $x$ contains
uncertainties as already mentioned.

\begin{figure}[bt]
\includegraphics[width=100mm]{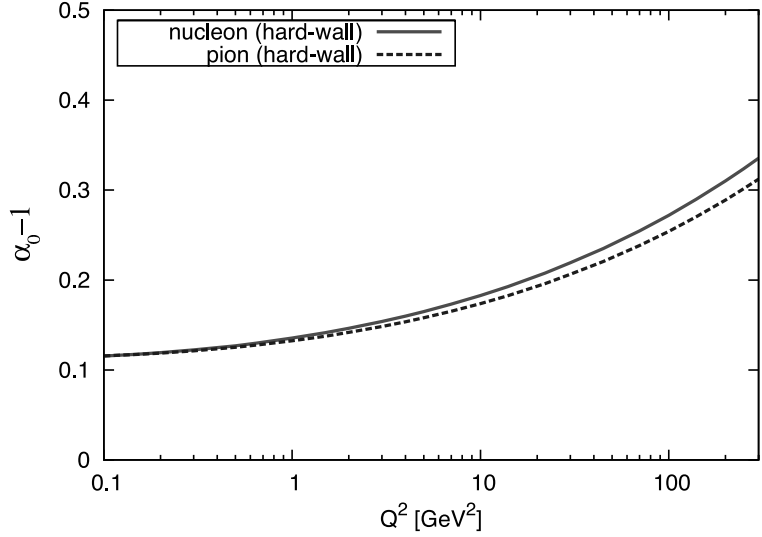}
\caption{ The effective Pomeron intercept $\alpha_0(Q^2)$ for the 
pion (dashed curve) is shown with the nucleon result (solid curve).  }%
\label{fig:pialpha}%
\end{figure}

The effective Pomeron intercept is also calculated for the pion 
case, shown in Fig.~\ref{fig:pialpha}.   
Both magnitude and $Q^2$-dependence are similar with the nucleon 
case, 
which may indicate the universality of the Pomeron intercept for
 various hadrons.

\begin{figure}[bt]
\includegraphics[width=65mm]{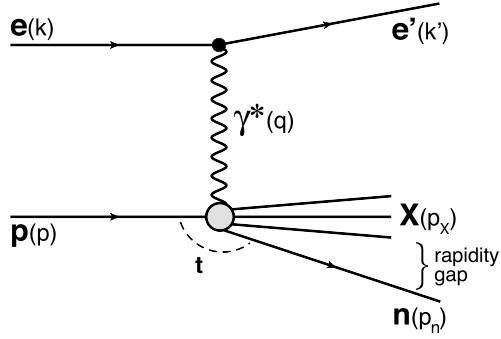}
\caption{The forward neutron production with the large rapidity gap 
by $\gamma^* p$ 
scattering.}%
\label{fig:fn}%
\end{figure}

\begin{figure}[bt]
\includegraphics[width=65mm]{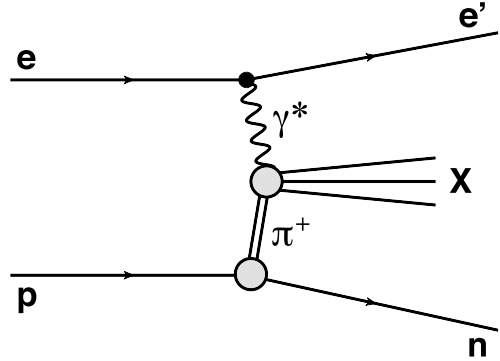}
\caption{The forward neutron production with the pion exchange model.}%
\label{fig:fnpi}%
\end{figure}

There also exist experimental information on the pion structure 
function from a forward neutron production in the semi-inclusive 
deep inelastic scattering off the nucleon~\cite{Aaron:2010ab} 
as illustrated in Fig.~\ref{fig:fn}.  
This process could be understood as a convolution of the 
deep inelastic lepton pion scattering cross section with the 
momentum distribution of the pion produced by the $\pi p n$ 
interaction 
in Fig.~\ref{fig:fnpi}, if there is a large rapidity gap between 
jets from the pion and the forward neutron.

The cross section of the semi-inclusive  $e+p \to e'+n+X$ process is 
assumed to be given by 
\begin{equation}
d\sigma \left( {ep \to e'nX} \right) = {f_{{\pi ^ + }/p}}\left( {{x_L},t} \right) \cdot d\sigma \left( {e{\pi ^ + } \to e'X} \right) \; ,
\end{equation}
where $f_{\pi ^ + /p}$ is the pion flux from the nucleon,  $x_L$ 
the longitudinal momentum fraction  carried by the leading 
neutron,   
and $t$ the 4-momentum transfer to the pion.  
The flux can be calculated with the standard pion-nucleon interaction in the infinite momentum frame~\cite{Holtmann:1994rs};
\begin{align}
{f_{{\pi ^ + }/p}}\left( {{x_L},t} \right) = \frac{1}{{2\pi }} \frac{{g_{p\pi n}^2}}{{4\pi }} \left( {1 - {x_L}} \right)\frac{{ - t}}{{{{\left( {m_\pi ^2 - t} \right)}^2}}} \exp \left( { - R_{\pi n}^2\frac{{m_\pi ^2 - t}}{{1 - {x_L}}}} \right) \; ,
\end{align}
where 
$m_\pi$ is the  pion mass, the coupling constant $g_{p\pi n}^2/4\pi =13.6$, 
and the exponential form factor with 
$R_{\pi n}=0.93 \ \mathrm{GeV}^{-1}$~\cite{Holtmann:1994rs}.  
Integrating ${f_{{\pi ^ + }/p}}$ over the momentum transfer, 
we obtain the pion flux from the nucleon as 
\begin{equation}
{\Gamma _\pi }\left( {{x_L}} \right) = \int_{{t_0}}^{{t_{\min }}} {{f_{{\pi ^ + }/p}}\left( {{x_L},t} \right)dt} \; ,
\label{piflux}
\end{equation}
where
\begin{align}
&{t_{\min }} =  - \left( {1 - {x_L}} \right)\left( {\frac{{m_n^2}}{{{x_L}}} - m_p^2} \right) \; ,  \\
&{t_0} =  - \frac{{{{\left( {p_T^{\max }} \right)}^2}}}{{{x_L}}} + {t_{\min }} \; .
\end{align}
We set 
$x_L = 0.73$ and $p_T^{\max } =0.2 \mbox{GeV} $ from HERA 
data~\cite{Aaron:2010ab}.   Resulting pion flux yields $0.133$, 
although this value contains non-negligible errors due to 
uncertainties of the measured $x_L$ and $p_T$.  
There may be additional contributions from the $\Delta$ mediated process, 
$p \to \Delta \pi$~\cite{Thomas:1998sc}, which we do not take into account in 
this work.

\begin{figure}[bt]
\includegraphics[width=130mm]{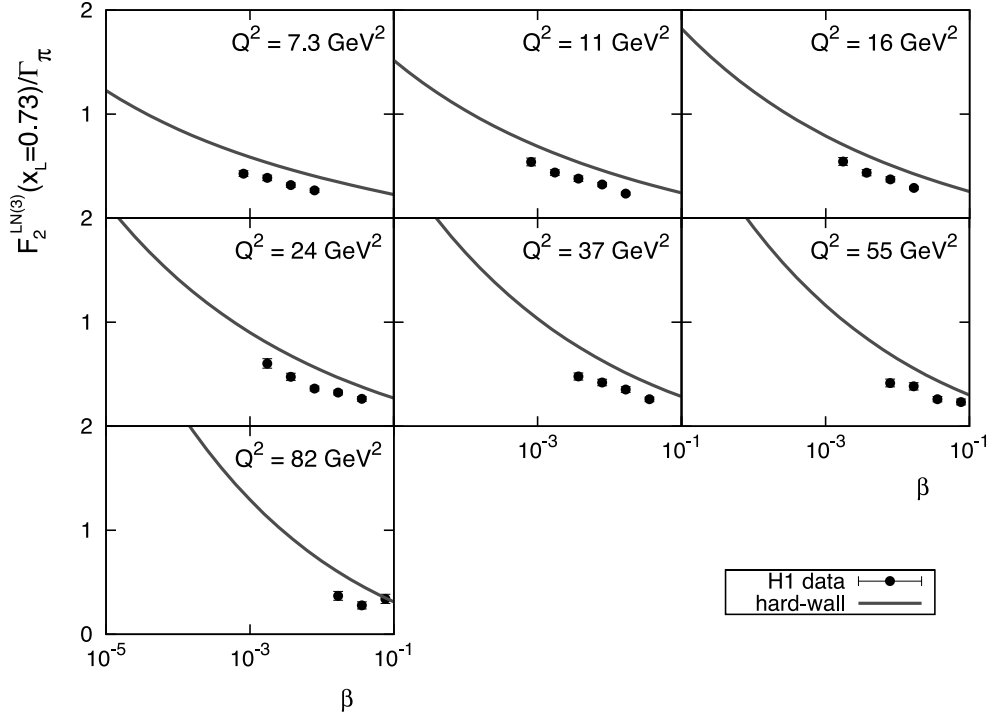}
\caption{The forward neutron production in the semi-inclusive 
DIS.    
Our calculation and the experimental data~\cite{Aaron:2010ab} are shown by the solid curves and circles, respectively.}
\label{fig:F3LN}
\end{figure}

Using our calculation for $F^\pi_2$, 
we compare the results with the experimental data in Fig.~\ref{fig:F3LN}.  
Because this analysis depends on the inputs of $x_L$ and $p_T$ 
in Eq.~(\ref{piflux}), 
absolute 
magnitudes of the calculations involve theoretical errors which may be 
about $30 \%$ at most.  
Hence, our results are consistent with the experimental data qualitatively.

\section{\label{sec:level6}Summary and Discussions}

We have calculated the nucleon and pion structure functions at the 
small-$x$ and 
studied the interplay between the soft and hard Pomerons 
in terms of the holographic 
QCD.   
The structure function at the small-$x$ is given by a convolution of the BPST 
Pomeron exchange 
kernel $\chi(s,z,z')$ with 
overlap functions of the virtual photon  $P_{13}( z,Q^2)$ and the 
target hadron $P_{24}(z')$ 
in the AdS space.   
We have emphasized the behavior of  $P_{13}(z)$ and $P_{24}(z')$ in the 
AdS space is a key to determine the value of the  
Pomeron intercept.  
Only when the peak position of $P_{13}(z,Q^2)$ in the $z$-space is away from 
the distribution of $P_{24}(z')$ in the $z'$-space, 
the resulting Pomeron intercept $\alpha_0$ is enhanced.  
This tendency is demonstrated by the super local (delta function) 
approximation for 
the overlap functions in Sec.~\ref{sec:level2}.

To be realistic, we have calculated the wave functions of the nucleon 
and the pion, which are the solutions of the classical action 
in the AdS space, and evaluated 
their couplings to the Pomeron.  
The Pomeron(graviton)-hadron-hadron coupling can be calculated by 
perturbing the metric in the classical action, and 
are essentially proportional to the 
square of the holographic wave functions in the AdS space.

As already known, the photon part 
$z P_{13}(z)$ has a flat distribution in the AdS $z$ space
 at $Q^2 \sim 0$, while 
it shows a 
sharp peak at $z=0$ for  $Q^2 \gg 1 \mbox{GeV}^2$.  
On the other hand, overlap functions $P_{24}(z')$ 
of the nucleon and the pion  are  shown to  be concentrated in the 
larger $z'$ region, near the hard-wall cutoff $z_0$.  
Such a behavior of the overlap functions are consistent with what 
we require to describe the transition between soft and hard Pomerons.

With these inputs 
we have calculated the nucleon structure function $F^p_2(x,Q^2)$ at the 
low-$x$.  The 
results fairly agree with the experimental data.  In particular, $Q^2$ dependence of 
the effective Pomeron intercept $\alpha_0 (Q^2)$ is consistent with the data, 
namely,  
the value of the Pomeron intercept $\alpha_0$ increases from the soft value 
$1.1$ to the hard one $1.3$  as 
the photon virtuality $Q^2$ increases from non-perturbative to the 
perturbative 
regions.

In general, our results are very similar with Ref.~\cite{Brower:2010wf} 
where {\em ad hoc}  
super local 
(delta function) approximation is used for the overlap functions.  
Nevertheless, our work is the first attempt to calculate the nucleon 
structure function at the small-$x$ with 
the holographic QCD in a
systematic way.  
The present approach enables us to calculate the scattering 
cross section of 
various hadrons without additional free parameters 
by evaluating their distributions in the AdS space.

We have newly calculated the pion structure function $F_2^\pi(x,Q^2)$ in 
this model.  
The resulting pion structure function at the small-$x$ is reduced by 
 about $30 \%$ in magnitude compared with the nucleon case.  
Unfortunately, 
we could not compare our calculations with the data, 
because the existing parameterizations of the pion structure function 
may already include large errors.  
However, the semi-inclusive DIS with a forward neutron production 
provides (model dependent) constraints on the pion structure function, 
if one assume the pion exchange model for this process.  
Our calculations are consistent with the data within theoretical 
uncertainties.

Because of the lack of experimental information on $F^\pi_2(x,Q^2)$, a 
simple relation between nucleon and pion structure functions 
is often used to estimate $F_2^\pi(x,Q^2)$ 
as~\cite{Nikolaev:1999qh,Chekanov:2002pf}, 
\begin{equation}
F_2^\pi (x,Q^2) \simeq \frac{2}{3} F_2^p 
\left( \frac{2}{3} x,Q^2 \right) \; .
\label{scaling}
\end{equation}
This relation simply originates  from the difference of the 
number of the valence quarks 
in the pion and the nucleon, although it is questionable whether 
or not this relation can be applied to the 
 small-$x$ region, at which the gluon dominates.

\begin{figure}[bt]
 \begin{center}
  \includegraphics[width=90mm]{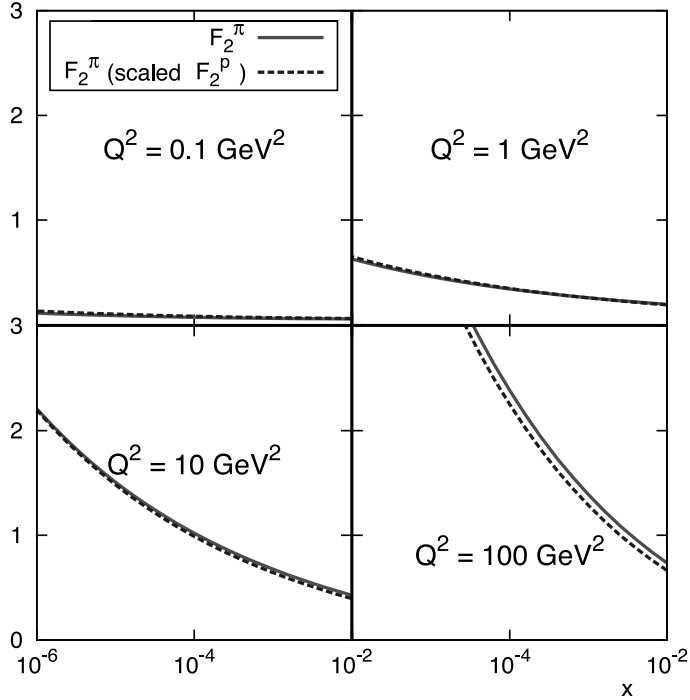}
 \end{center}
 \caption{Calculated $F_2^\pi$ (solid curve) from holography and the result with the scaling relation Eq.~(\ref{scaling}) (dashed curve).}
 \label{fig:F2_NSCALEDvsMP}
\end{figure}

We show in 
Fig.~\ref{fig:F2_NSCALEDvsMP} the calculated pion $F_2^\pi$ by the solid 
curve (LHS of Eq.~(\ref{scaling})) and 
a result calculated by the scaling relation (\ref{scaling}) with the 
input of the calculated nucleon $F_2^p$(RHS of (\ref{scaling})).   
It is interesting to see this relation 
is satisfied almost perfectly, 
although we cannot explain why it holds  within our framework.

It is our strong desire to extend the present approach to the 
hadron-hadron scattering.  In this case, 
the scattering amplitude of hadrons $i$ and $j$ is given by
\begin{align}
{\cal A} (s,t) = 2is \int d^2b \, e^{i \bm{q} \cdot \bm{b}} \int dzdz' P_i(z) P_j(z') \{ 1-e^{i \chi (s,b,z,z')} \} \; .
\end{align}
If $i$ and $j$ are the same hadron, $P_i(z)$ and $ P_j(z')$ have the same distribution in $z$ and $z'$ spaces, respectively.  According to our results for DIS, the Pomeron intercept $\alpha_0$ becomes 
small (soft) in case the distribution of the incident and target particles 
are similar.  
Therefore, we expect the hadron scattering of the identical particles 
can be 
described by the soft Pomeron,
which is consistent with the phenomenologies.  
Even if we consider the hadron scattering of the different species,
shapes of $P_i(z)$ and $ P_j(z')$ in the AdS space are not so different.  
Thus, we also expect the hadron scattering 
of different particles 
to be described by the soft Pomeron exchange.  These studies are 
now in progress.

Another possible extension of this work is to consider the 
diffractive photo- and 
lepto-production of neutral vector mesons, $\rho$, $\phi$, J$/\psi$, $\Upsilon$~\cite{Ivanov:2004ax}.  
Experimental data of these processes also clearly show  the scale dependence of the 
Pomeron intercept.  
In this case, the quark mass as well as the photon virtuality play roles of the 
hard scale, 
In order to apply our framework, it is necessary to evaluate the photon-Pomeron-vector 
meson coupling in the AdS space, which is   under considerations.

\begin{acknowledgments}
A.W.~acknowledges G.~Ogawa for useful comments on the holographic description 
of the nucleon.  
We also thank all the members of the quark/hadron group in Tokyo University of
Science for useful conversations.   
\end{acknowledgments}


\begin{thebibliography}{30}                                                                       

\bibitem{FS} See, {\em e.~g.~}
J. R. Forshaw and D. A. Ross, "Quantum Chromodynamics and the Pomeron", (Cambridge University Press, London, 1997).
%

\bibitem{DL} 
A.~Donnachie and P.~V.~Landshoff, Phys. Lett. {\bf B470}, 243 (1999).


\bibitem{Breitweg:1998dz}
  J.~Breitweg {\it et al.}  [ZEUS Collaboration],
  Eur.\ Phys.\ J.\  C {\bf 7}, 609 (1999)
  [arXiv:hep-ex/9809005].



\bibitem{BFKL} 
E.A.~Kuraev, L.N.~Lipatov and V.S.~Fadin, Sov.~Phys.~JETP {\bf 45}
 (1977), 199; 
Ya.Ya.~Balitsky and L.N. Lipatov, Sov.~J.~Nucl.~Phys.~{\bf 28} (1978), 22.



\bibitem{Ivanov:2004ax} For a review, 
  I.~P.~Ivanov, N.~N.~Nikolaev and A.~A.~Savin,
  Phys.\ Part.\ Nucl.\  {\bf 37}, 1 (2006)
  [hep-ph/0501034], and references therein.



\bibitem{Maldacena:1997re} 
  J.~M.~Maldacena,
  Adv.\ Theor.\ Math.\ Phys.\  {\bf 2}, 231 (1998)
  [Int.\ J.\ Theor.\ Phys.\  {\bf 38}, 1113 (1999)]
  [hep-th/9711200].

\bibitem{Gubser:1998bc} 
  S.~S.~Gubser, I.~R.~Klebanov and A.~M.~Polyakov,
  Phys.\ Lett.\ B {\bf 428}, 105 (1998)
  [hep-th/9802109].

\bibitem{Witten:1998qj} 
  E.~Witten,
  Adv.\ Theor.\ Math.\ Phys.\  {\bf 2}, 253 (1998)
  [hep-th/9802150].

\bibitem{review} 
  O.~Aharony, S.~S.~Gubser, J.~M.~Maldacena, H.~Ooguri and Y.~Oz,
  Phys.\ Rept.\  {\bf 323}, 183 (2000)
  [hep-th/9905111].

\bibitem{SonStephanov} 
  D.~T.~Son and M.~A.~Stephanov,
  Phys.\ Rev.\ D {\bf 69}, 065020 (2004)
  [hep-ph/0304182].


\bibitem{Erlich:2005qh} 
  J.~Erlich, E.~Katz, D.~T.~Son and M.~A.~Stephanov,
  Phys.\ Rev.\ Lett.\  {\bf 95}, 261602 (2005)
  [hep-ph/0501128].


\bibitem{Polchinski:2001tt} 
  J.~Polchinski and M.~J.~Strassler,
  Phys.\ Rev.\ Lett.\  {\bf 88}, 031601 (2002)
  [hep-th/0109174].



\bibitem{review_meson} 
  J.~Erdmenger, N.~Evans, I.~Kirsch and E.~Threlfall,
  Eur.\ Phys.\ J.\ A {\bf 35}, 81 (2008)
  [arXiv:0711.4467 [hep-th]].

\bibitem{review_bottom} 
For reviews,   J.~Erlich,
  PoS CONFINEMENT {\bf 8}, 032 (2008)
  [arXiv:0812.4976 [hep-ph]], 
  J.~Erlich,
  Int.\ J.\ Mod.\ Phys.\ A {\bf 25}, 411 (2010)
  [arXiv:0908.0312 [hep-ph]].


\bibitem{Brower:2006ea} 
  R.~C.~Brower, J.~Polchinski, M.~J.~Strassler and C.~-I.~Tan,
  JHEP {\bf 0712}, 005 (2007)
  [hep-th/0603115].

\bibitem{Brower:2007qh} 
  R.~C.~Brower, M.~J.~Strassler and C.~-I.~Tan,
  JHEP {\bf 0903}, 050 (2009)
  [arXiv:0707.2408 [hep-th]].

\bibitem{Brower:2007xg} 
  R.~C.~Brower, M.~J.~Strassler and C.~-I.~Tan,
  JHEP {\bf 0903}, 092 (2009)
  [arXiv:0710.4378 [hep-th]].

\bibitem{Brower:2010wf} 
  R.~C.~Brower, M.~Djuric, I.~Sarcevic and C.~-I.~Tan,
  JHEP {\bf 1011}, 051 (2010)
  [arXiv:1007.2259 [hep-ph]].




\bibitem{Aaron:2010ab} 
  F.~D.~Aaron {\it et al.}  [H1 Collaboration],
  Eur.\ Phys.\ J.\ C {\bf 68}, 381 (2010)
  [arXiv:1001.0532 [hep-ex]].



\bibitem{Polchinski:2002jw} 
  J.~Polchinski and M.~J.~Strassler,
  JHEP {\bf 0305}, 012 (2003)
  [hep-th/0209211].


\bibitem{Hatta:2007he}
  Y.~Hatta, E.~Iancu and A.~H.~Mueller,
  JHEP {\bf 0801}, 026 (2008)
  [arXiv:0710.2148 [hep-th]].

\bibitem{Levin:2009vj} 
  E.~Levin, J.~Miller, B.~Z.~Kopeliovich and I.~Schmidt,
  JHEP {\bf 0902}, 048 (2009)
  [arXiv:0811.3586 [hep-ph]].


\bibitem{Henningson:1998cd} 
  M.~Henningson and K.~Sfetsos,
  Phys.\ Lett.\ B {\bf 431}, 63 (1998)
  [hep-th/9803251].

\bibitem{Muck:1998iz} 
  W.~Muck and K.~S.~Viswanathan,
  Phys.\ Rev.\ D {\bf 58}, 106006 (1998)
  [hep-th/9805145].

\bibitem{Contino:2004vy} 
  R.~Contino and A.~Pomarol,
  JHEP {\bf 0411}, 058 (2004)
  [hep-th/0406257].

\bibitem{Hong:2006ta} 
  D.~K.~Hong, T.~Inami and H.~-U.~Yee,
  Phys.\ Lett.\ B {\bf 646}, 165 (2007)
  [hep-ph/0609270].



\bibitem{Kwee:2007dd} 
  H.~J.~Kwee and R.~F.~Lebed,
  JHEP {\bf 0801}, 027 (2008)
  [arXiv:0708.4054 [hep-ph]].


\bibitem{Grigoryan:2007wn} 
  H.~R.~Grigoryan and A.~V.~Radyushkin,
  Phys.\ Rev.\ D {\bf 76}, 115007 (2007)
  [arXiv:0709.0500 [hep-ph]].


\bibitem{Grigoryan:2008up} 
  H.~R.~Grigoryan and A.~V.~Radyushkin,
  Phys.\ Rev.\ D {\bf 77}, 115024 (2008)
  [arXiv:0803.1143 [hep-ph]].


\bibitem{Abidin:2009hr} 
  Z.~Abidin and C.~E.~Carlson,
  Phys.\ Rev.\ D {\bf 79}, 115003 (2009)
  [arXiv:0903.4818 [hep-ph]].

\bibitem{Abidin:2008hn} 
  Z.~Abidin and C.~E.~Carlson,
  Phys.\ Rev.\ D {\bf 77}, 115021 (2008)
  [arXiv:0804.0214 [hep-ph]].

\bibitem{Grigoryan:2008cc} 
  H.~R.~Grigoryan and A.~V.~Radyushkin,
  Phys.\ Rev.\ D {\bf 78}, 115008 (2008)
  [arXiv:0808.1243 [hep-ph]].


\bibitem{Aaron:2009aa}
  F.~D.~Aaron {\it et al.}  [H1 and ZEUS Collaboration],
  JHEP {\bf 1001}, 109 (2010)
  [arXiv:0911.0884 [hep-ex]].


\bibitem{Sutton:1991ay} 
  P.~J.~Sutton, A.~D.~Martin, R.~G.~Roberts and W.~J.~Stirling,
  Phys.\ Rev.\ D {\bf 45}, 2349 (1992).


\bibitem{Gluck:1999xe} 
  M.~Gluck, E.~Reya and I.~Schienbein,
  Eur.\ Phys.\ J.\ C {\bf 10}, 313 (1999)
  [hep-ph/9903288].


\bibitem{Aicher:2010cb}
M.~Aicher, A.~Schafer and W.~Vogelsang,
  Phys.\ Rev.\ Lett.\  {\bf 105}, 252003 (2010)
  [arXiv:1009.2481 [hep-ph]].


\bibitem{Holtmann:1994rs} 
  H.~Holtmann, G.~Levman, N.~N.~Nikolaev, A.~Szczurek and J.~Speth,
  Phys.\ Lett.\ B {\bf 338}, 363 (1994).



\bibitem{Thomas:1998sc}
 A.~W.~Thomas and C.~Boros,
 Eur.\ Phys.\ J.\ C {\bf 9}, 267 (1999)
 [hep-ph/9812264].




  

\bibitem{Nikolaev:1999qh} 
  N.~N.~Nikolaev, J.~Speth and V.~R.~Zoller,
  Phys.\ Lett.\ B {\bf 473}, 157 (2000)
  [hep-ph/9911433].



\bibitem{Chekanov:2002pf} 
  S.~Chekanov {\it et al.}  [ZEUS Collaboration],
  Nucl.\ Phys.\ B {\bf 637}, 3 (2002)
  [hep-ex/0205076].



\end{thebibliography}
\end{document}